# Realizing giant valley polarization effect based on monolayer altermagnets[*]

XIE Weifeng[1], WANG Libo[1], XU Xiong[2], YUE Yunliang[3], XIA Huayan[1], HE Longhui[1], WANG Hui[2]

1.School of Microelectronics and Physics, Hunan University of Technology and Business, Changsha 410205, China

2.School of Physics, Central South University, Changsha 410083, China

3.College of Information Engineering, Yangzhou University, Yangzhou 225127, China

**Abstract**

Stable and remarkable valley polarization effect is the key to utilizing valley degree of freedom in valleytronic devices. According to first-principles calculations and symmetry analysis, we reveal that valley polarization effect in monolayer $V_2Se_2O$ altermagnet is correlated with the net magnetic moment between magnetic V atoms under uniaxial strain, thereby proposing two strategies for achieving giant valley polarization effect. Firstly, substituting one V atom in $V_2Se_2O$ with Cr to construct a ferrimagnetic monolayer $VCrSe_2O$ enhances the net magnetic moment between magnetic atoms, thereby realizing a giant valley polarization effect. Applying uniaxial strain along either the *a*-axis or *b*-axis significantly increases the value of valley polarization, which exhibits a nearly linear relationship with the net magnetic moments between the magnetic atoms. Secondly, constructing a van der Waals heterostructure composed of $V_2Se_2O$ and α-SnO monolayers breaks mirror symmetry, thereby inducing a net magnetic moment, which in turn causes a remarkable valley polarization effect. Compressing the interlayer distance of the heterostructure can increase the net magnetic moment between V atoms, then enhancing the value of valley polarization to nearly 400 meV. This work reveals that valley polarization in monolayer altermagnet is correlated with the net magnetic moment between magnetic atoms. Finally, we propose two strategies to achieve giant valley polarization based on monolayer altermagnets, providing theoretical guidance for the potential applications of ferrimagnetic monolayers and altermagnet-based heterostructures in valleytronics.

**Keywords:** valley degree of freedom, monolayer altermagnets, valley polarization, ferrimagnetic, van der Waals heterostructure

## 1. Introduction

With the development of modern electronic devices towards miniaturization and integration, the inevitable quantum effect of traditional semiconductor devices due to miniaturization, as well as the high energy consumption and difficult heat dissipation caused by charge transfer, have become increasingly prominent. Similar to the traditional charge and spin degrees of freedom, the valley degree of freedom can be used as a new type of information carrier, which can encode information by using the occupied states of electrons in different energy valleys, so that the storage, transmission and processing of data do not involve charge flow, thus opening up a new way for the development of semiconductor devices and valleytronic devices in the post-Moore era[1,2].

The valley polarization effect is the premise of the application of the valley degree of freedom. The valley polarization can be expressed as the energy difference between different valleys, that is,

$$\Delta E_{v(c)} = E_{v(c)}^{K} - E_{v(c)}^{K'}, \tag{1}$$

Where v and c denote the conduction and valence bands, respectively; $K$ and $K'$ mark the two valley points, respectively. Exploring valley materials with giant valley polarization effect and exploring the formation mechanism of valley polarization have always been the key to the application of valley degree of freedom[3]. Nonmagnetic transition metal disulfides ($MoS_2$ and $WSe_2$) have attracted much attention due to their broken inversion symmetry and large spin-orbital coupling (SOC) effect, which leads to large spin splitting at different valley points[4,5]. However, the protection of time-reversal symmetry makes the energy degenerate between different valleys[6]. At present, some external means, such as external magnetic field[6-9], magnetic atom doping[10,11], magnetic proximity effect[12-15], circularly polarized light excitation[16-18] and so on, are used to eliminate the intervalley degeneracy, thus forming the valley polarization feature which is beneficial to the application of the valley degree of freedom. However, the implementation conditions of these external means are harsh, which makes people look forward to finding an intrinsic valley material with large valley polarization. In 2016, Tong et al.[19] discovered the phenomenon of spontaneous valley polarization formed by perpendicular magnetization under SOC in ferromagnetic transition metal disulfide 2H-VSe$_2$, and first proposed the concept of ferrovalley property. The Hamiltonian of the ferrovalley material can be expressed as

$$\hat{H}(k) = \hat{H}_0(k) + \hat{H}_{\text{SOC}}(k) + \hat{H}_{\text{ex}}(k), \qquad (2)$$

Where $\hat{H}_{\text{SOC}}(k)$ denotes the Hamiltonian of the SOC term; $\hat{H}_{\text{ex}}(k)$ represents the Hamiltonian of the intrinsic magnetic exchange interaction. ferrovalley materials have intrinsically broken space and time reversal symmetry, and the anomalous valley Hall effect occurs under appropriate carrier doping and in-plane electric field, which is more conducive to the use of valley degree of freedom[19-24]. So far, a series of two-dimensional ferromagnetic and antiferromagnetic materials have been proved to have ferrovalley behavior[25-29]. However, in general, ferrovalley materials need to rely on SOC to form valley polarization, and the valley polarization value induced by SOC is not very large, most of which are below 100 meV[30,31]. Finding valley materials with large valley polarization is still a research hotspot in the field of valleytronics.

Recently, a new type of collinear magnetic material, altermagnet[32-36], which combines the advantages of ferromagnet and antiferromagnet, has attracted much attention. The altermagnet has an antiparallel magnetic order compensated similar to an antiferromagnet, and its unique rotational symmetry, mirror symmetry of different sublattices, and broken time reversal symmetry make the altermagnet with valleys have a spin-valley locking phenomenon[37] caused by crystal symmetry rather than time reversal symmetry at the valley point, which is equivalent to the SOC induced spin splitting phenomenon in non-magnetic valley materials under the protection of time reversal symmetry. It is worth noting that the giant spin splitting present in the altermagnet can be an order of magnitude larger than the SOC-induced spin splitting[38]. At present, many monolayer altermagnetic semiconductors and valley-dependent physical properties have been theoretically predicted[37,39-47]. It is shown that most monolayer altermagnetic semiconductors have a uniaxial strain-induced non-SOC valley polarization effect, and the valley polarization value increases nearly linearly with the uniaxial strain. Under – 5% uniaxial strain, a significant valley polarization (>100 meV) can be produced, which is called the piezovalley effect[37]. The piezovalley effect effect in altermagnets breaks the precondition that SOC should be considered to produce valley polarization in conventional ferrovalley materials, and the valley polarization effect can occur by breaking the corresponding crystal symmetry by applying uniaxial strain[37,39-45]. Then in addition to uniaxial strain, we naturally think of whether there are other strategies to break the crystal symmetry in altermagnets?

In this paper, the mirror symmetry between different magnetic atoms in altermagnets is broken by replacing magnetic atoms and constructing Van der Waals heterojunctions. Through first-principles calculations, it is found that the uniaxial strain changes the net magnetic moment between magnetic atoms in $V_2Se_2O$ monolayer, accompanied by the change of valley polarization. Based on the correlation between the valley polarization and the net magnetic moment, a Cr atom is used to replace one V atom in the $V_2Se_2O$ monolayer, resulting in a large net magnetic moment between the magnetic atoms. It is found that the $VCrSe_2O$ monolayer is a ferrimagnetic semiconductor, and a giant valley polarization can occur without considering the SOC, which is similar to the piezovalley effect in monolayer altermagnets[37,39]. When uniaxial strain is applied to the *a* axis and *b* axis of $VCrSe_2O$ monolayer, respectively, it is found that stretching the *a* axis and compressing the *b* axis can significantly increase the magnitude of valley polarization, accompanied by the change of the net magnetic moment between magnetic atoms. Another strategy to change the net magnetic moment between V atoms in $V_2Se_2O$ monolayer is to construct van der Waals heterojunction. We find that in the heterojunction constructed by $V_2Se_2O$ monolayer and α-SnO monolayer, the stacking structure that breaks the mirror symmetry can change the net magnetic moment between V atoms, resulting in significant valley polarization. By controlling the interlayer distance, it is found that the compression of the interlayer distance can increase the net magnetic moment and the valley polarization at the same time, and the valley polarization reaches nearly 400 meV when the compression is 0.5 Å. In this work, two strategies to achieve giant valley polarization based on monolayer altermagnets are proposed, which provide theoretical guidance for the application of ferrimagnetic monolayer and heterojunction materials based on altermagnets in the field of valleytronics.

## 2. Calculation method

The VASP (Vienna *ab initio* simulation package) software package for[48] calculation based on density functional theory was used in this work. The projector augmented plane wave method[49] and the Perdew-Burke-Ernzerhof (PBE) exchange-correlation functional based on the generalized gradient approximation are used[50]. A plane wave cutoff energy of 560 eV is used in the calculation, and a vacuum layer of more than 15 Å perpendicular to the two-dimensional plane is used to avoid the influence of interlayer interaction on the calculation results. The energy convergence criteria for the structure optimization and self-consistent calculations were unified as $1 \times 10^{-7}$ eV, and the convergence threshold for the force on each atom was unified as 0.005 eV/Å. For the selection of *K* grid, the Monkhorst-Pack method with 25×25×1 centered at *Γ* point is used to calculate $V_2Se_2O$ monolayer. A *K* point grid of 24 × 25×1 was used to calculate the $VCrSe_2O$ monolayer. In the calculation of $V_2Se_2O$/SnO heterojunction, a *K*-point mesh of 25×25×1 is used. In addition, in order to further consider the strong correlation effect of *d* electrons of V and Cr atoms, the PBE +*U* method[51] is used in the calculation, in which the $U_{eff}$ of V is set to 4 eV[39], and the $U_{eff}$ of Cr is set to 3.55 eV[41,42]. The interlayer van der Waals interaction in $V_2Se_2O$/SnO heterojunction was also modified by DFT-D3 method[52].

## 3. Results and Discussion

$V_2Se_2O$ monolayer is a typical two-dimensional altermagnet[37] with *P4/mmm* space group, and the crystal structure is shown in Fig. 1(a). Since the $V_2Se_2O$ monolayer is tetragonal with in-plane lattice constants *a* =*b* = 4.044 Å, the different sublattices have vertical mirror symmetry along the diagonal direction, as shown by the mirror $M_\phi$ in Fig. 1(a) and Fig. 1(b). In addition to the mirror symmetry, the $V_2Se_2O$ monolayer has a unique rotational symmetry, which is also a distinctive feature that distinguishes altermagnets from antiferromagnets[32,33]. It can be seen from the local structure of the two V atoms of Fig. 1(c) that V is located at the center of the distorted octahedron and the two V atoms have mirror symmetry. The net magnetic moment $\Delta M(V_1 - V_2)$ between V atoms is defined as

$$\Delta M (V_1 - V_2) = |M_{V_1}| - |M_{V_2}|. \tag{3}$$

First-principles calculations show that the $\Delta M(V_1 - V_2)$ of $V_2Se_2O$ monolayer is zero. From the band structure of the Fig. 1(d), it can be seen that there is a spin-split band characteristic along the high symmetry path of *M-X-Γ-Y-M*, and the whole system is a direct band gap semiconductor with a band gap value of $E_g$ = 0.738 eV. It has been theoretically demonstrated that $V_2Se_2O$ monolayer has a significant valley polarization effect induced by uniaxial strain, and the valley polarization does not need to consider the SOC effect, only by applying a uniaxial strain (Fig. 1(e)) in the direction of *a* axis (or *b* axis), the valley polarization phenomenon can occur, which is called the piezovalley effect[37]. We uniformly define the calculation formula of the valley polarization value $\Delta E_{c(v)}$ of the monolayer altermagnet with tetragonal crystal system:

$$\Delta E_{c(v)} = E_{c(v)}(X) - E_{c(v)}(Y), \tag{4}$$

Where c and v denote the lowermost conduction band (LCB) and the uppermost valence band (UVB), respectively, while X and Y mark two valley points along the high-symmetry path, as illustrated in the inset of Figure 1(d). First-principles calculations indicate that applying approximately 5% compressive strain along the $a$-axis induces a valley polarization of 85.203 meV in the UVB of the $V_2Se_2O$ monolayer. Interestingly, the calculations reveal that although the total magnetic moment of the $V_2Se_2O$ monolayer does not vary with uniaxial strain, the uniaxial strain does alter $\Delta M(V_1 - V_2)$, causing $\Delta M(V_1 - V_2)$ to change linearly with the applied uniaxial strain, as shown by the blue line in Figure 1(f). Consequently, the valley polarization of the UVB exhibits a trend with $\Delta M(V_1 - V_2)$ that is consistent with its trend with uniaxial strain, as depicted in the inset of Figure 1(f). From this phenomenon, we can infer that the ability of uniaxial strain to modulate valley polarization is reflected in the variation of the net magnetic moment between magnetic atoms in an altermagnetic material with strain. This provides a new approach for tuning valley polarization in monolayer altermagnetic materials, namely, by modifying the net magnetic moment between magnetic atoms through external means.

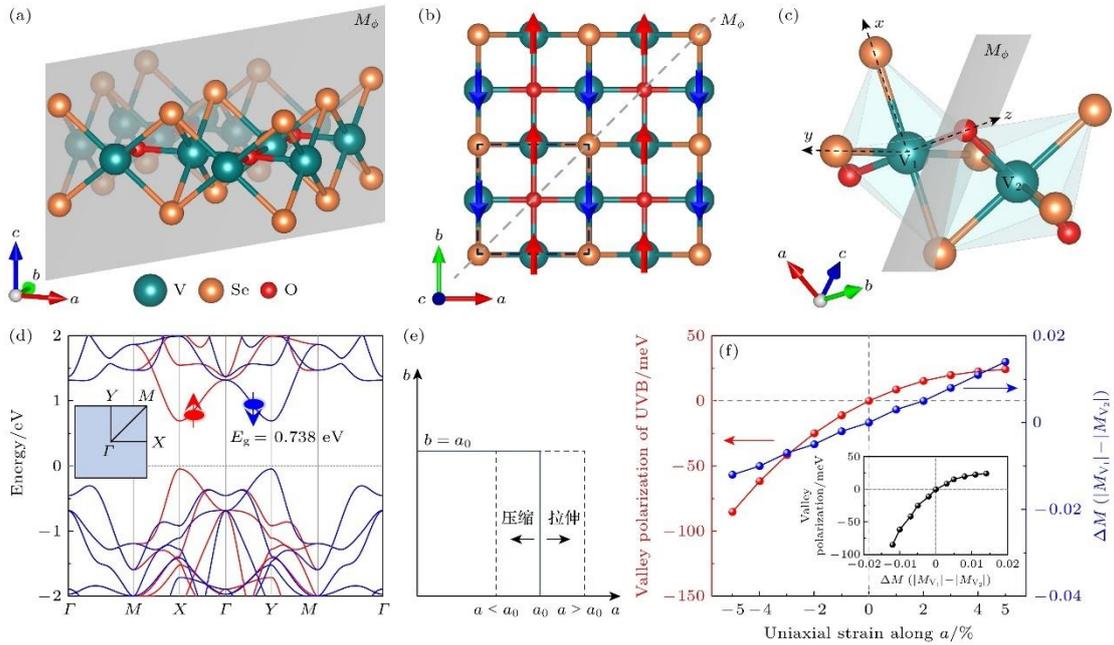

**Figure 1.** Geometric and band structures of the $V_2Se_2O$ monolayer, and the relationship among valley polarization, net magnetic moment, and uniaxial strain along the $a$-axis direction: (a) Crystal structure of monolayer $V_2Se_2O$, where the gray planes in the vertical direction represent the mirror symmetry planes $M_\phi$ of different sublattices along the diagonal direction; (b) top view of the monolayer $V_2Se_2O$, where the black dashed box indicates the unit cell, and the gray dashed lines

represent the mirror planes $M_\phi$ along the diagonal direction; (c) local octahedral structure of two V atoms, where *xyz* represent the local octahedral coordinates of one V atom; (d) spin-resolved band structure calculated by PBE + $U$ ($U$ = 4 eV), where the inset shows the high-symmetry points and paths in the first Brillouin zone; (e) schematic illustration of uniaxial strain along the *a*-axis; (f) valley polarization of UVB and the net magnetic moment $\Delta M(V_1 - V_2)$ between V atoms as a function of *a*-axis strain, where the inset shows the variation of valley polarization with $\Delta M(V_1 - V_2)$.

Previous studies have shown that the piezovalley effect[37] is caused by the mirror symmetry breaking of the $V_2Se_2O$ monolayer due to uniaxial strain. We find that the regulation of valley polarization by uniaxial strain is indirectly reflected in the change of $\Delta M(V_1 - V_2)$. Uniaxial strain, as a means of regulation, can change the $\Delta M(V_1 - V_2)$, accompanied by the formation of valley polarization. It is natural to think that one V atom in the $V_2Se_2O$ monolayer can be replaced by an element replacement strategy, which can not only break the mirror symmetry of different sublattices, but also make a large net magnetic moment between magnetic atoms. Based on the fact that the magnitude of valley polarization of $V_2Se_2O$ monolayer increases with the increase of $\Delta M(V_1 - V_2)$ under uniaxial compressive strain, we infer that the system can produce significant valley polarization effect by replacing magnetic atoms in a 1∶1 ratio. We have tested the element substitution between the same group and different groups with V atom, and found that the $VCrSe_2O$ monolayer formed by replacing one V atom in the $V_2Se_2O$ monolayer by Cr atom in a 1:1 ratio has a stable geometric structure. Firstly, the formation energy $E_f$ of $VCrSe_2O$ monolayer is calculated by first-principles calculations. The formation energy is calculated by

$$E_f = (E_{VCrSe_2O} - \mu_V - \mu_{Cr} - 2 \times \mu_{Se} - \mu_O)/5,$$

Where $E_{VCrSe_2O}$ is the energy of monolayer $VCrSe_2O$, $\mu_V$ is the chemical potential of V in bcc single crystal of V, $\mu_{Cr}$ is the chemical potential of Cr in bcc single crystal of Cr, $\mu_{Se}$ is that of Se in trigonal Se single crystal, and $\mu_O$ is that of O in $O_2$ molecule. The calculated formation energy of $VCrSe_2O$ is – 1.677 eV, which is nearly 4 times that of $CrWI_6$ monolayer (– 0.474) and 8 times that of $CrWGe_2Te_6$ monolayer (– 0.216 eV)[53]. The negative value of the calculated formation energy indicates that the reaction is exothermic, which is more conducive to the stable formation of the compound in experiment. At the same time, we list six different site substitution structures in which V and Cr atoms may exist in a 1:1 ratio in the 2×2 supercell of the $VCrSe_2O$ monolayer, as shown in Fig. 2(a). At the same time, the relative energies of the six structures were calculated, as shown in Fig. 2(b). It can be seen that the P1 substitution structure corresponds to the lowest energy, and compared with other structures, the energy difference of each formula unit is at least 120 meV, which is nearly 4 times larger than the energy difference of different substitution sites of $CrWI_6$ and $CrWGe_2Te_6$ monolayers[53]. Fig. 2(c) is the top view and side view of $VCrSe_2O$ monolayer P1structure. Due to the lattice constant $a \neq b$, the mirror symmetry $M_\phi$ is broken, which can theoretically produce significant valley polarization without considering the SOC effect.

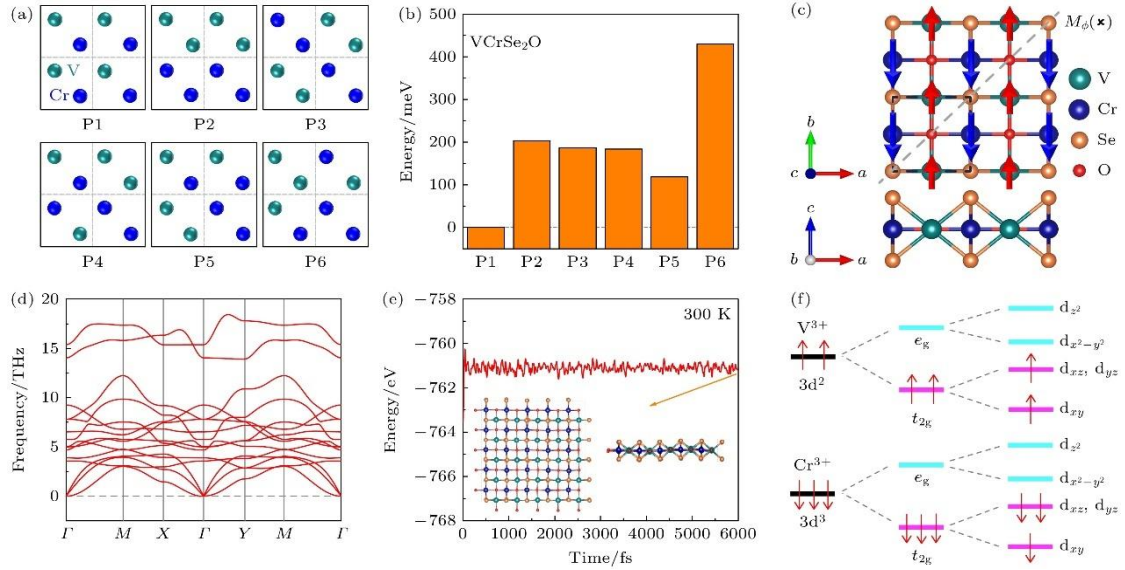

**Figure 2.** Stability analysis and schematic electronic configuration of the monolayer $VCrSe_2O$: (a) Different substitution configurations in a 2×2 supercell of the $VCrSe_2O$ monolayer, where V and Cr atoms are located at different sites with a 1:1 ratio, only V and Cr atoms are labeled; (b) relative energies of the different substitution configurations, where P1–P6 correspond to the six structures in panel (a), with the energy of the P1 structure taken as the reference; (c) top and side views of the P1 structure of monolayer $VCrSe_2O$, where the gray dashed lines represent the vertical mirror planes $M_\phi$ of different sublattices along the diagonal direction, here the mirror symmetry is broken; (d) phonon spectrum of the monolayer $VCrSe_2O$; (e) AIMD simulation of the monolayer $VCrSe_2O$ at 300 K; (f) distribution of energy levels and electrons of the magnetic atoms V and Cr in the local distorted octahedra.

Firstly, the stability of the P1structure of $VCrSe_2O$ monolayer is analyzed. The phonon spectrum without imaginary frequency in the Fig. 2(d) shows that the $VCrSe_2O$ monolayer has dynamic stability. The Fig. 2(e) is the *ab initio* molecular dynamics (AIMD) simulation at 300 K. From the figure, it can be seen that the $VCrSe_2O$ monolayer can still maintain the structure without deformation within 6 *ps*, so it has thermodynamic stability. The above analysis makes it possible to synthesize $VCrSe_2O$ monolayers experimentally. The results of first-principles calculation show that the local magnetic moment of V atom is 2.093 $\mu_B$, and the local magnetic moment of Cr atom is – 3.263 $\mu_B$, and the two local magnetic moments are in opposite directions. The whole system has a net magnetic moment of – 1 $\mu_B$, and the energy of this ferrimagnetic structure is 388.427 meV lower than that of the ferromagnetic structure, indicating that the magnetic ground state of the VCrSe$_2$O monolayer is ferrimagnetic. From the local magnetic moments of V atoms near 2 $\mu_B$ and Cr atoms near – 3 $\mu_B$, it can be seen that the VCrSe$_2$O monolayer is similar to the V$_2$Se$_2$O monolayer, V atoms and Cr atoms lose three electrons each, the electronic structure of V atoms changes from $3d^34s^2$ to $3d^2$, and the electronic structure of Cr atoms changes from $3d^54s^1$ to $3d^3$. Since the Cr atom replaces one V atom in the V$_2$Se$_2$O monolayer, both the V and Cr atoms in the VCrSe$_2$O monolayer are in a local octahedron, similar to Fig. 1(c). The regular octahedron structure can split the five $d$ orbitals into a two-fold degenerate $e_g$ and a lower energy three-fold degenerate $t_{2g}$. In the local octahedron formed by four Se atoms and two O atoms around the V atom, the V-O bond is 0.655 Å shorter than the V-Se bond. In the local octahedron formed by four Se atoms and two O atoms around the Cr atom, the Cr-O bond is 0.576 Å shorter than the Cr-Se bond. The $xy$ plane is constructed by V atom and four Se atoms, and the direction of V-O bond is the $z$ axis, which is similar to the dotted local coordinate in Fig. 1(c). Therefore, the local distorted octahedron formed by V and Cr is equivalent to the regular octahedron compressed along the $z$ axis. The final result of the crystal field results in the splitting of the two-fold degenerate $e_g$ into $d_{z^2}$ and the lower energy $d_{x^2-y^2}$ orbital, and the three-fold degenerate $t_{2g}$ into the two-fold degenerate $d_{xz}$, $d_{yz}$, and the lower energy $d_{xy}$ orbital, as shown in Fig. 2(f). For the V atom, the two spin-up $d$ electrons fill the $d_{xy}$ and $d_{xz}$ ($d_{yz}$) orbitals according to the Pauli exclusion principle and the Hund rule, forming a high-spin state and contributing a local magnetic moment of 2 $\mu_B$. For the Cr atom, three spin-down $d$-electrons fill the $d_{xy}$, $d_{xz}$ and $d_{yz}$ orbitals, forming a high-spin state and contributing a local magnetic moment of – 3 $\mu_B$. The final result is that the VCrSe$_2$O monolayer is a ferrimagnet with a total magnetic moment of – 1 $\mu_B$.

The spin density maps of VCrSe$_2$O monolayer are shown in Fig. 3(a) and Fig. 3(b). The spin densities of V atoms and Cr atoms are no longer mirror symmetry and rotational symmetry, but have opposite signs, and the spin density of Cr atoms is larger than that of V atoms, which is in line with the first-principles calculations. The spin-resolved band of VCrSe$_2$O monolayer is shown in the Fig. 3(c). It is obvious that VCrSe$_2$O monolayer is a ferrimagnetic semiconductor with an indirect band gap of 632.137 meV. Moreover, there is a giant valley polarization at the high symmetry points $X$ and $Y$, and the valley polarization value reaches 161.824 meV, which is nearly twice the valley polarization value of V$_2$Se$_2$O monolayer under 5% uniaxial compressive strain. The perturbation of SOC on the energy band is calculated by Fig. 3(d). Compared with the giant valley polarization induced by symmetry breaking, the influence of SOC on the valley polarization is very small. In the subsequent regulation of valley polarization, we ignore the influence of SOC on valley polarization. According to the band structure of element decomposition (Fig. 3(e) and Fig. 3(f)), the electronic states at $X$ are contributed by spin-up V atoms and spin-up Se atoms, while the electronic states at $Y$ are contributed by spin-down Cr atoms and spin-down Se atoms, so the net magnetic moment between V atoms and Cr atoms can affect the magnitude of valley polarization.

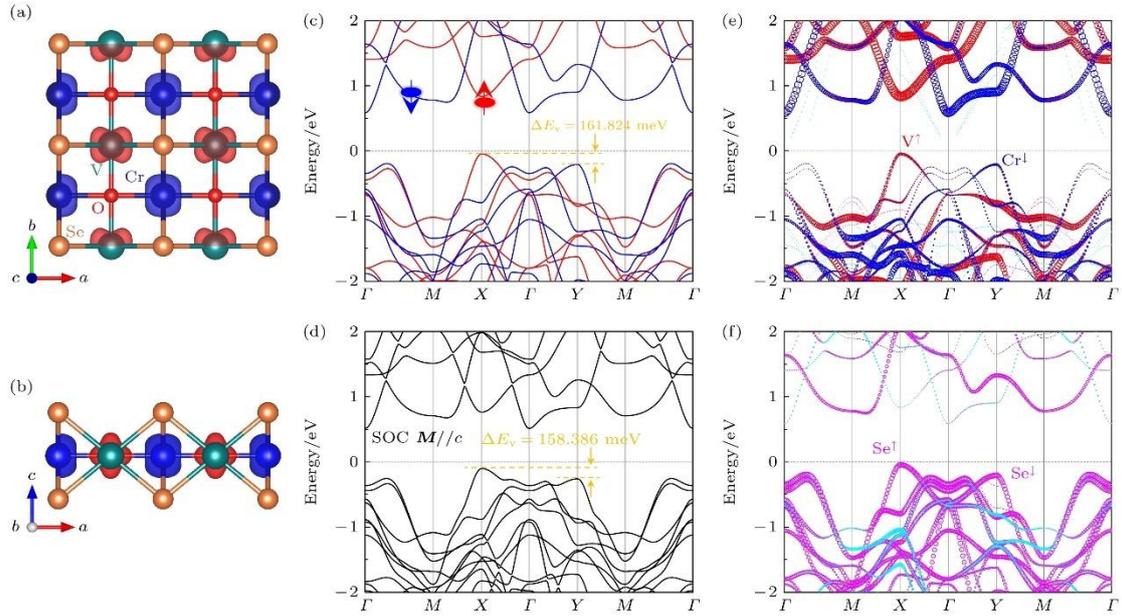

**Figure 3.** Electronic structure and orbital distribution of the VCrSe$_2$O monolayer: (a) Top view and (b) side view of spin density of monolayer VCrSe$_2$O, where the isosurface is set to 0.05 $e$/Å$^3$; (c) spin-resolved band structure and (d) SOC-perturbated band structure, where the magnetization $M$ is parallel to $c$-axis and the value of valley polarization of the UVB is marked in yellow; (e) contributions of V and Cr atoms on band structure, where the size of the circles indicates the weight; (f) contributions of Se atoms on band structure.

Similar to V$_2$Se$_2$O monolayer, uniaxial strain can still be used as a control method to control the valley polarization in VCrSe$_2$O monolayer. First, the net magnetic moment $\Delta M$(V−Cr) between V and Cr atoms in VCrSe$_2$O monolayer is defined as

$$\Delta M (\text{V} - \text{Cr}) = |M_\text{V}| - |M_\text{Cr}|. \tag{5}$$

The changes of valley polarization and $\Delta M$(V−Cr) of VCrSe$_2$O monolayer along the $a$ axis and $b$ axis under -5% — 5% uniaxial strain were calculated by first-principles calculations, and the calculation results are shown in Fig. 4(a) and Fig. 4(d). It can be seen from the red line in the Fig. 4(a) that the compressive strain along *a* axis decreases the valley polarization, while the tensile strain increases the valley polarization slowly at first and then tends to be stable after 3%. As can be seen from the blue line in the Fig. 4(a), the ΔM(V−Cr) tends to be nearly linear with strain, except for tensile strains after 3%. It can be concluded from Fig. 4(a) that the valley polarization of UVB in VCrSe$_2$O monolayer increases with the size of Δ$M$(V−Cr), which is similar to that in V$_2$Se$_2$O monolayer under uniaxial strain. We have plotted the band structures for the cases of 5% compression and 5% extension along the *a* axis, respectively, as shown in Fig. 4(b) and Fig. 4(c). 5% compressive strain reduces the valley polarization to 85.798 meV, and 5% tensile strain increases the valley polarization to 177.755 meV, but the maximum point of the valence band is not at the high symmetry points of *X* and *Y*, which is not conducive to the application of valley polarization.

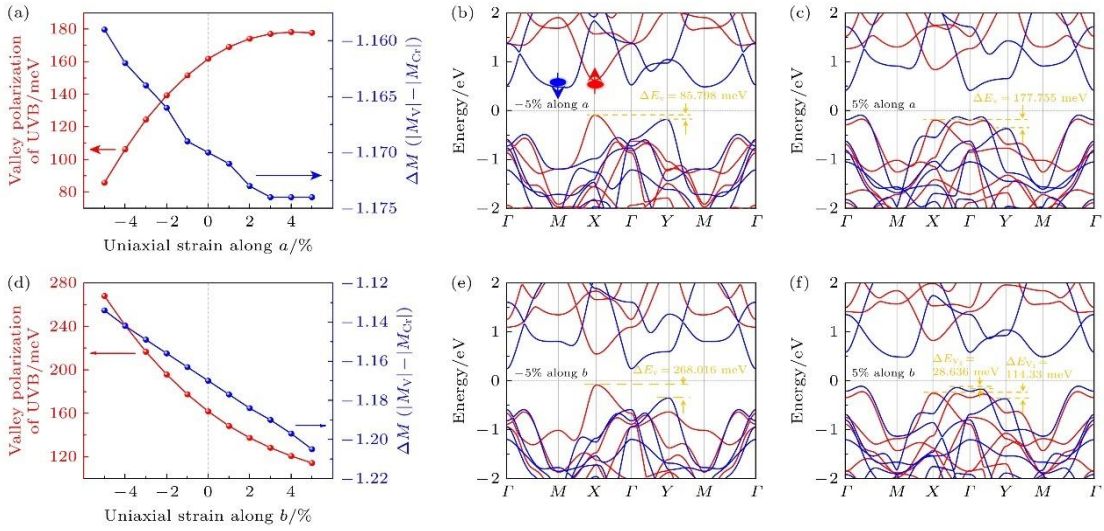

**Figure 4.** Valley polarization, net magnetic moment and band structure of VCrSe$_2$O monolayer under uniaxial strain along *a*-axis and *b*-axis directions: (a) and (d) Variation of valley polarization of the UVB and Δ$M$(V − Cr) with uniaxial strain along the *a*-axis and *b*-axis, respectively, for the monolayer VCrSe$_2$O; (b) and (e) the spin-resolved band structures under –5% compressive strain along the *a*-axis and *b*-axis, respectively; (c) and (f) the spin-resolved band structures under 5% tensile strain along the *a*-axis and *b*-axis, respectively. On the band structures, the values of valley polarization corresponding to the relevant uniaxial strains are marked in yellow.

The Fig. 4(d) shows the effect of the *b* axis strain on the valley polarization and Δ$M$(V−Cr) of UVB. It can be seen that the compressive strain can significantly increase the valley polarization, while the tensile strain decreases the valley polarization. The relationship between Δ$M$ (V−Cr) and uniaxial strain is still nearly linear. Unlike the case of the *a* axis strain, the value of the *b* axis strain-induced valley polarization decreases with the increase of the Δ$M$ (V−Cr). This can be explained by the change in the relative ratio between the *a* axis and the *b* axis. In the V$_2$Se$_2$O monolayer, uniaxial compression or tension increases the relative ratio between the *a* axis and the *b* axis, and correspondingly increases the valley polarization. For the VCrSe$_2$O monolayer, the lattice constants of the *a* and *b* axes are 4.099 Å and 4.010 Å, respectively, and the stretching of the *a* axis increases the relative ratio between the *a* and *b* axes, resulting in a larger valley polarization. Similarly, when compressed along the *b* axis, the relative ratio between the *a* axis and the *b* axis increases, thereby inducing a larger valley polarization. The spin-polarized band structure corresponding to a compressive strain of 5% along the *b* axis is shown in Fig. 4(e), where the valley polarization reaches 268.016 meV. The band structure for 5% tensile strain is shown in Fig. 4(f), where the valley polarization is reduced to 114.33 meV. Interestingly, the 5% tensile strain along the *b* axis makes the two valleys on the *Γ-X* and *Γ-Y* high symmetry paths in the valence band more prominent, and the calculated results show that the energy difference between the valleys reaches 28.636 meV. Therefore, under a tensile strain of 5% along the *b* axis, a double valley polarization effect appears in the VCrSe$_2$O monolayer, which is obviously different from the conventional valley polarization system, thus making the VCrSe$_2$O monolayer have special application value in the field of valleytronics.

In order to break the mirror symmetry of different sublattices in altermagnets and make a net magnetic moment between different magnetic atoms, another feasible strategy is to construct a van der Waals heterojunction[54], which changes the local environment of different magnetic atoms by controlling the stacking mode between layers, accompanied by the breaking of mirror symmetry, resulting in a significant valley polarization effect. It is found that α-SnO monolayer has the same tetragonal[55] as V$_2$Se$_2$O monolayer, and the lattice mismatch of their unit cells is about 5%, so it can form a stable heterostructure. By controlling the position of the upper Sn atom in α-SnO, we constructed three stacking structures with high symmetry. The upper Sn atom is located under the O atom of the V$_2$Se$_2$O monolayer, and the resulting stacking structure is named VSH1 (van der Waals stacked heterojunctions 1); The upper Sn atom is located below the Se atom and is named VSH2; the upper Sn atom is located below the V atom and is named VSH3. The top and side views of the ground states of these three stacking structures are shown in Fig. 5. VSH1 is found to be the most stable stacking structure. In contrast, the energy of VSH2 is 19.878 meV higher than that of VSH1, and the energy of VSH3 is similar to that of VSH2. We have calculated the band structures and layer-decomposed band structures for the three stacking cases, as shown in Fig. 6. It is found that there is no valley polarization in VSH1 and VSH2, mainly because the two stacking modes do not destroy the mirror symmetry of different sublattices, as shown by Fig. 6(a) and Fig. 6(b). Interestingly, through the layer-decomposed band structure (Fig. 6(d),(e)), it is found that the band structure entirely contributed by the α-SnO layer exhibits a small spin splitting on the high-symmetry path *M-X-Γ-Y-M*. We can conclude that when the non-magnetic system α-SnO monolayer and the staggered magnet V$_2$Se$_2$O monolayer form a heterojunction, due to the magnetic neighbor effect of the V$_2$Se$_2$O monolayer, the non-magnetic system α-SnO layer appears spin splitting, thus having altermagnetism.

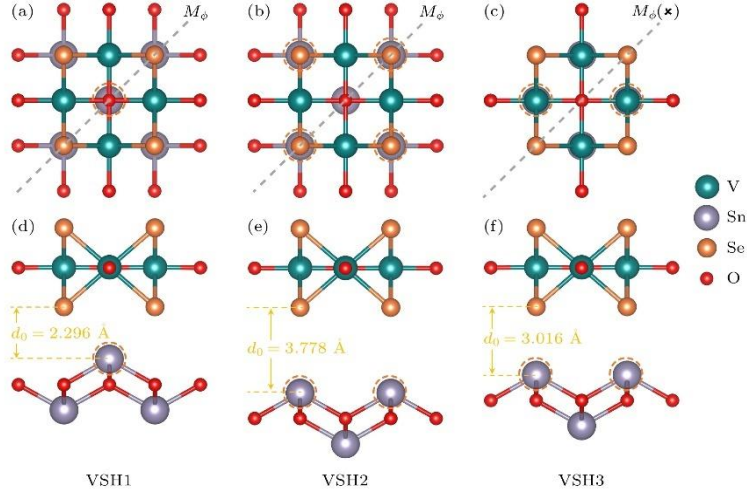

**Figure 5.** Three stacked configurations of V$_2$Se$_2$O/SnO heterojunctions: (a) and (d), (b) and (e), (c) and (f) correspond to the top and side views of three stacked structures of the V$_2$Se$_2$O/SnO heterojunction, respectively, where the dashed circles indicate the upper Sn atoms in the α-SnO layer and the interlayer distances of the heterojunctions under different stacked models are marked in yellow.

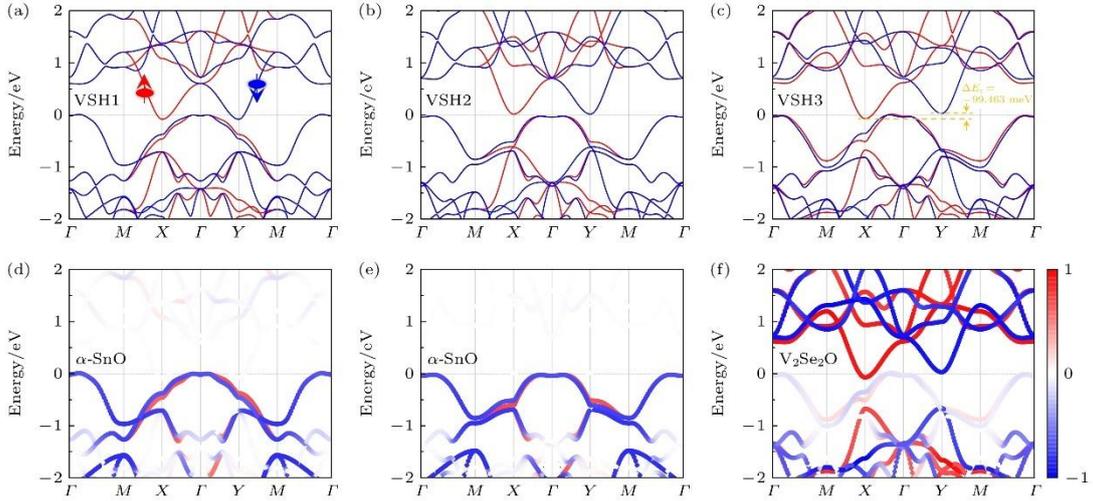

Figure 6. Layer-resolved band structures of V$_2$Se$_2$O/SnO heterojunctions corresponding to three stacked configurations: (a)–(c) The spin-resolved band structures of the V$_2$Se$_2$O/SnO heterojunction under three different stacked structures, respectively, where the values of valley polarization of the LCB in the VSH3 are marked in yellow; (d)–(f) correspond to the layer-resolved band structures for the three stacked structures, respectively, where the chromatic bands represent the weights of projected spin-up (red) and spin-down (blue) electrons on the bands in the relevant layers.

VSH3 is different from VSH1 and VSH2 due to the different local environments of the two V atoms, which leads to the difference of lattice constants in the directions of $a$ axis and $b$ axis, $a$ = 3.975 Å, $b$ = 3.989 Å. The mirror symmetry breaking must make the $\Delta M$(V1−V2) non-zero, resulting in significant valley polarization. As shown in the Fig. 6(c) and Fig. 6(f), the band structure from first-principles calculations and the layer-decomposed band structure show that in VSH3, due to the special stacking mode of α-SnO and $V_2Se_2O$ monolayers, there is a significant valley polarization in the band contributed by $V_2Se_2O$ monolayers, and the valley polarization of LCB reaches 99.463 meV.

In theory, the electronic state of the heterojunction can be effectively controlled by controlling the interlayer distance of the heterojunction, which is reflected in the change of the band structure[56]. In this paper, the variation of valley polarization and $\Delta M$(V1−V2) with interlayer distance in VSH3 is calculated by first-principles calculations, as shown in the Fig. 7(a). The interlayer distance of VSH3 in equilibrium was found to be $d_0$ = 3.016 Å (Fig. 5(f)). Decreasing the interlayer distance can significantly increase the valley polarization of the $V_2Se_2O$ layer. At $d_0$ - 0.3 Å, the valley polarization reaches 220.804 meV. At this time, the bands of the $V_2Se_2O$ layer and the SnO layer do not overlap, as shown in Fig. 7(b). At $d_0$ - 0.5 Å, the magnitude of valley polarization reaches a giant 379.189 meV. Although there is an overlap between the energy bands of the $V_2Se_2O$ layer and the SnO layer, it is found from the band structure of the layer decomposition that the energy band at the valley point is mainly contributed by the $V_2Se_2O$ layer, so the valley polarization comes from the upper interlaced magnet $V_2Se_2O$ layer. Increasing the interlayer distance decreases the magnitude of the valley polarization, which approaches zero at $d_0$ + 0.5 Å. The blue line of Fig. 7(a) shows that $\Delta M$(V1−V2) decreases with increasing interlayer distance, and $\Delta M$(V1−V2) tends to zero at $d_0$ + 0.5 Å. Since the magnitude of valley polarization decreases with decreasing $\Delta M$(V1−V2), the value of valley polarization approaches zero at zero net magnetic moment, as shown by the red line of Fig. 7(a). This phenomenon also further confirms the relationship between valley polarization and the net magnetic moment between magnetic atoms in altermagnets.

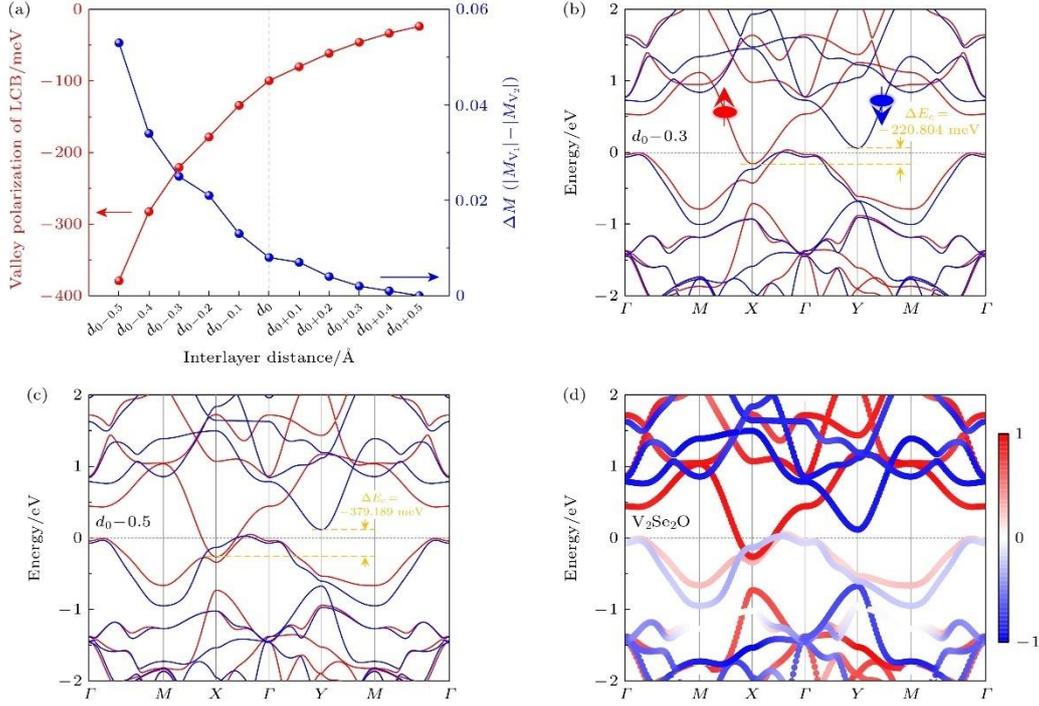

**Figure 7.** Valley polarization and net magnetic moment versus interlayer distance, and layer-resolved band structures in VSH3: (a) Variation of the valley polarization of the LCB and $\Delta M(V_1 - V_2)$ with interlayer distance in VSH3; (b) and (c) the spin-resolved band structures under interlayer compression of 0.3 and 0.5 Å, respectively, where the values of valley polarization of the LCB are marked in yellow; (d) contribution of band from the $V_2Se_2O$ layer under interlayer compression of 0.5 Å, where the chromatic bands represent the weights of projected spin-up (red) and spin-down (blue) electrons on the bands.

## 4. Conclusion

In the altermagnet $V_2Se_2O$ monolayer, it is found that the uniaxial strain can changes the magnitude of the net magnetic moment between V atoms, accompanied by a change in valley polarization. Based on the correlation between the net magnetic moment and the valley polarization, two strategies to achieve giant valley polarization are proposed on the basis of monolayer altermagnets, namely, the replacement of magnetic atoms and the construction of Van der Waals heterojunctions. The ferrimagnetic monolayer $VCrSe_2O$ is formed by replacing one V atom in the $V_2Se_2O$ monolayer by a Cr atom, and the large net magnetic moment causes a giant valley polarization of the $VCrSe_2O$. In addition, the *a*-axis and *b*-axis strain can significantly enhance the valley polarization of $VCrSe_2O$ monolayer, accompanied by a change in the net magnetic moment. By constructing $V_2Se_2O$/SnO van der Waals heterojunction, it is found that the mirror symmetry-breaking stacking mode can lead to the appearance of net magnetic moment between V atoms and significant valley polarization of the heterojunction. A giant valley polarization of nearly 400 meV can be achieved with a compressed interlayer distance of 0.5 Å, accompanied by an increase in the net magnetic moment between V atoms. In this work, the correlation between the net magnetic moment and valley polarization of magnetic atoms in monolayer altermagnets is found, and two strategies to achieve giant valley polarization are proposed, which lays a foundation for the research of ferrimagnetic monolayer and heterojunction materials based on altermagnets in the field of valleytronics.